\documentclass[useAMS,usenatbib,fleqn]{mnras}
\usepackage{subfigure}
\usepackage{graphicx}
\usepackage{times}
\usepackage{tabularx}
\usepackage{natbib}
\usepackage{url} 
\usepackage{xcolor}
\usepackage{caption}
\usepackage{amsmath}
\usepackage{float}
\usepackage{multirow}
\usepackage{morefloats}
\usepackage{amssymb}

\newcommand{\dd}{\mathrm{d}}
\newcommand{\Mpc}{\mathrm{Mpc}}
\newcommand{\Msun}{\mathrm{M}_{\odot}}

\newcommand{\kpc}{\mathrm{kpc}}

\newcommand{\K}{\mathrm{K}}
\newcommand{\Gyr}{\mathrm{Gyr}}
\newcommand{\Myr}{\mathrm{Myr}}

\newcommand{\fc}{F_{\mathrm{comp}}}

\newcommand{\fst}{F_{\mathrm{stretch}}}
\newcommand{\fb}{F_{\mathrm{baro}}}
\newcommand{\fdiss}{F_{\mathrm{diss}}}
\newcommand{\fadv}{F_{\mathrm{adv}}}

\newcommand{\vvec}{\mathbf{v}}
\newcommand{\pc}{\mathrm{pc}}

\definecolor{myred}{rgb}{1,0,0} 
\definecolor{myblue}{rgb}{0,0,1}
\definecolor{mygreen}{rgb}{0,1,0}

\def\stacksymbols #1#2#3#4{\def\theguybelow{#2}
        \def\verticalposition{\lower#3pt}
        \def\spacingwithinsymbol{\baselineskip0pt\lineskip#4pt}
        \mathrel{\mathpalette\intermediary#1}}
\def\intermediary #1#2{\verticalposition\vbox{\spacingwithinsymbol
        \everycr={}\tabskip0pt
        \halign{$\mathsurround0pt#1\hfil##\hfil$\crcr#2\crcr
                \theguybelow\crcr}}}

\begin{document}
\label{firstpage}
\pagerange{\pageref{firstpage}--\pageref{lastpage}}
 
 \title[Black Hole Weather Storms]
 {Unraveling Baroclinicity in Black Hole Weather Storms}
 \author[Wittor \& Gaspari]{Denis Wittor$^{1,2}$\thanks{E-mail: dwittor@ha.uni-hamburg.de} 
 \& Massimo Gaspari$^{3}$\thanks{E-mail: mgaspari@astro.princeton.edu}\\
 $^{1}$ Hamburger Sternwarte, Gojenbergsweg 112, 21029 Hamburg, Germany \\
 $^{2}$ Dipartimento di Fisica e Astronomia, Universita di Bologna, Via Gobetti 93/2, I-40122 Bologna, Italy \\
 $^{3}$ Department of Astrophysical Sciences, Princeton University, 4 Ivy Lane, Princeton, NJ 08544-1001, USA \\
  \vspace{-1.0cm}
  }
 \maketitle

 \vspace{-0.2cm}
 \begin{abstract} 
    In the intracluster, intragroup, and circumgalactic medium (ICM, IGrM, CGM), turbulence plays a vital role in the self-regulated feedback and feeding cycle of central supermassive black holes (SMBHs). Here we continue our systematic dissection of the turbulent `weather' in high-resolution hydrodynamical simulations of feedback driven by active galactic nuclei (AGN). In non-barotropic and stratified atmospheres, baroclinicity is expected to generate fresh turbulence via misaligned gradients of density and pressure -- such as in cyclonic storms on Earth.
    In this work, we dissect for the first time baroclinicity and its components in the astrophysical halo weather. Over the macro-scale galaxy cluster, baroclinicity tends to be dynamically subdominant for the enstrophy amplification. However, at and below the meso scale near the SMBH ($r<10$ kpc; $t<20$ Myr), baroclinicity is important to seed the initial enstrophy during active periods of AGN jet feedback. 
    We find that baroclinicity shows stronger correlation with the density rather than pressure gradients. 
    Despite the density-pressure gradient misalignment being often below $45^{\circ}$, their amplitudes boosted by mechanical AGN feedback are sufficient to enable key enstrophy/turbulence generation.
    Our study provides a novel step forward in understanding astrophysical atmospheres toward a unified \texttt{BlackHoleWeather} framework, akin to the complexity of Earth's weather.
 \end{abstract}

 \begin{keywords} galaxies: clusters: intracluster medium - hydrodynamics - turbulence - (galaxies:) quasars: supermassive black holes - galaxies: active - methods: numerical
  \vspace{-0.5cm}
 \end{keywords}
 
 \section{Introduction} \label{s:intro}
 Turbulence is one of the key ingredients in the self-regulated feeding and feedback cycle of active galactic nuclei (\citealt{2007_McNamara_Nulsen,Fabian:2012, Eckert:2021}, for reviews). The central supermassive black holes (SMBH) is expected to grow recurrently via Chaotic Cold Accretion (CCA) \citep[e.g.][and references therein]{Gaspari:2013_cca,Gaspari:2017_cca,Voit:2018}, i.e.~the condensation of cold clouds and warm filaments out of the turbulent atmospheres that rain on the central SMBH of the host ICM, IGrM, or CGM. The AGN feedback, which is triggered in response, then quenches the feeding and cooling flows by releasing substantial mechanical energy into the ambient medium via collimated jets and ultrafast outflows. This feedback injects turbulence on scales of 5\,-\,100 kpc around the SMBH, thus shaping the evolution of galaxies, groups and clusters of galaxies \citep[e.g.][for a review]{Gaspari:2020}.
 
 In \citet{2020MNRAS.498.4983W} (hereafter Paper 1), we studied the evolution of turbulence injected by AGN feedback by following the enstrophy in hydrodynamical simulations. Local enstrophy, which can be defined as the amplitude of vorticity, $\epsilon \equiv \frac{1}{2} |\boldsymbol{\nabla} \times \vvec|^2$, is a well-known proxy for solenoidal turbulence \citep[e.g.][]{2017MNRAS.464...210V}. The evolution of enstrophy is determined by advection, compression, stretching, dissipation and baroclinicity. 
 For Eulerian hydrodynamics, the equation that describes the evolution of enstrophy is
 \vspace{-0.5cm}
  \begin{align}
 \left(\frac{\dd \epsilon}{\dd t}\right)_{\mathrm{Euler}} &= \fb + \fadv + \fc + \fst + \fdiss \label{eq::enst_euler} \\
   \fb &= \frac{\boldsymbol{\omega}}{\rho^2} \cdot ( \boldsymbol{\nabla}  \rho \times \boldsymbol{\nabla}  P)
  = \boldsymbol{\omega}\,\frac{P}{\rho} \cdot \left( \frac{\boldsymbol{\nabla} \rho} {\rho} \times \frac{\boldsymbol{\nabla}  P} {P}\right)   \label{eq:Fbaro} %\\
  \end{align}
  \begin{align}
   \quad \quad \fadv &= - \boldsymbol{\nabla} \cdot (\vvec \epsilon) \label{eq:Fadv} \\
   \quad \quad \fc &= -\epsilon \boldsymbol{\nabla}  \cdot \vvec  \label{eq:Fcomp}\\
   \quad \quad \fst &= 2\epsilon (\boldsymbol{\hat{\omega}} \cdot \boldsymbol{\nabla}  ) \vvec \cdot \boldsymbol{\hat{\omega}}  \label{eq:Fstretch}\\
   \quad \quad \fdiss &= \nu \boldsymbol{\omega} \cdot \left[{\nabla}^2 \boldsymbol{\omega} + \boldsymbol{\nabla}  \times
  \left((1/\rho) \boldsymbol{\nabla}  \rho \cdot \mathbf{S}\right)\right] \label{eq:Fdiss}
 \end{align}
 where $\vvec$ is the velocity, $\boldsymbol{\omega}=\boldsymbol{\nabla}  \times \vvec$ is the vorticity (with $\boldsymbol{\hat{\omega}}$ its unit vector), $\mathbf{S}$ the strain tensor, $\nu$ the %kinematic
 viscosity (which is here negligible), $\rho$ and $P$ are the gas density and pressure. The above equations show that only baroclinicity can generate net enstrophy, while all the other terms require the presence of a minimum amount of $\varepsilon$ to be non-zero. For extended discussions on the different terms, we refer the interested reader to \citet{Porter:2015} and \citet{2017MNRAS.471.3212W}.

 In Paper 1, we studied the evolution of enstrophy and its dynamical terms in a high-resolution hydrodynamical simulation of AGN feedback leveraging both Eulerian and Lagrangian diagnostics. Over the macro-scales, compression and stretching determine the enstrophy evolution and, hence, of the turbulent motions in the diffuse gaseous medium, such as the ICM. Advection was found to be rarely relevant. At first sight, throughout most of the volume, baroclinicity seemed to remain subdominant compared with the other terms. However, we recently found that baroclinicity can become dynamically important in very localized patches at the micro- and meso-scale level near the SMBH.
 Therefore, in this Letter, we dissect the properties of the baroclinic term in magnified depth. 
 
 Baroclinicity generates enstrophy in non-barotropic and stratified atmospheres. For an adiabatic equation of state, as used here, enstrophy is generated due to misaligned gradients of pressure and density. 
 In meteorology, baroclinicity is a crucial term to be assessed, since it is the initial cause for the frequent formation of midlatitude-belt cyclones (e.g.~\citealt{Houze:2014}), which have a profound effect on Earth's weather. 
 At the same time, they are clearly not identical systems. The cyclonic nature of Earth's storms has a different long-term amplification mode. 
 In astrophysical atmospheres (ICM/IGrM/CGM), the vorticity increases mainly due to the misalignment behind curved shocks driven by the jet or within a multiphase cooling flow. 
 On the other hand, Earth's atmosphere has substantially faster rotation, with the Coriolis force mainly amplifying vorticity perpendicular to the gravity field.
 
 In sum, as part of \texttt{BlackHoleWeather} program (PI: Gaspari), we aim here to make a further step toward a comprehensive modeling of astrophysical atmospheres around SMBH, akin to Earth's weather systems.
 
 \vspace{-0.5cm}
 \section{Simulations}
 In this work, we use the same simulation setup that was first presented in \citep{Gaspari:2012b} and later analysed in Paper 1 \citep{2020MNRAS.498.4983W}. Here, we briefly summarise the main properties of the simulation and we refer the interested reader to such references for further details (e.g., for more in-depth discussions of the self-regulated AGN feeding and feedback processes).
 
 The simulation was carried out with the adaptive-mesh-refinement astrophysical code \texttt{Flash4} \citep[][]{Fryxell:2000}; by using the three-dimensional hydrodynamics equations it simulated a typical brightest cluster galaxy (BCG) within a typical cool-core cluster halo. The initial halo models the observations of Abell\,1795, with a virial mass of $M_{\mathrm{vir}} \sim 10^{15} \ \Msun$ and a virial temperature of $T_{\mathrm{vir}} \sim 10^7 \ \K$. The simulations include radiative cooling $\propto n^2 \Lambda$ (where $n$ is the gas number density and $\Lambda$ the cooling function; \citealt{Sutherland:1993}). This induces the loss of temperature/pressure from $T \sim 5 \times 10^7  \ \K$ to $10^4 \ \K$ in the gaseous halo and the related formation of condensed multiphase gas. The AGN feedback counteracts the radiative cooling, which would otherwise develop a catastrophic cooling flow in an unbalanced regime. In the nuclear region, the SMBH quenches the cooling by injecting kinetic energy and momentum via recurrent jets and ultrafast outflows, which further generate macro-scale shocks and buoyant bubbles. The macro heating mechanism affects a radius of $r \sim 100 \ \kpc$ around the SMBH. 

 The simulation domain covers a volume of size $(1.3 \ \Mpc)^3$, which is sampled with 10 levels of concentric static meshes, each refined by a factor of 2 toward the SMBH. %(SMR) 
 The maximum level has a resolution of $\Delta x \approx 300 \ \pc$.  
 Here, we focus on the $(170 \ \kpc)^3$ cluster core volume, which includes the central BCG and related CGM/ICM.
 Of particular relevance for the baroclinc study will be the inner sub-region, $r < 20 \ \kpc$.
 Furthermore, we analyse a period of $100 \ \Myr$, with a temporal resolution of $\Delta t = 0.1 \ \Myr$, taken out of the total $5\ \Gyr$ evolution. 
 The analysed period is of key interest for our study, as it covers a dozen typical strong and weak AGN outburst. We checked that it robustly samples the characteristic recurrent evolution of the AGN feedback cycle. 

 \vspace{-0.5cm}
 \section{Results}

 \begin{figure}
     \centering
     \includegraphics[width = 0.44\textwidth]{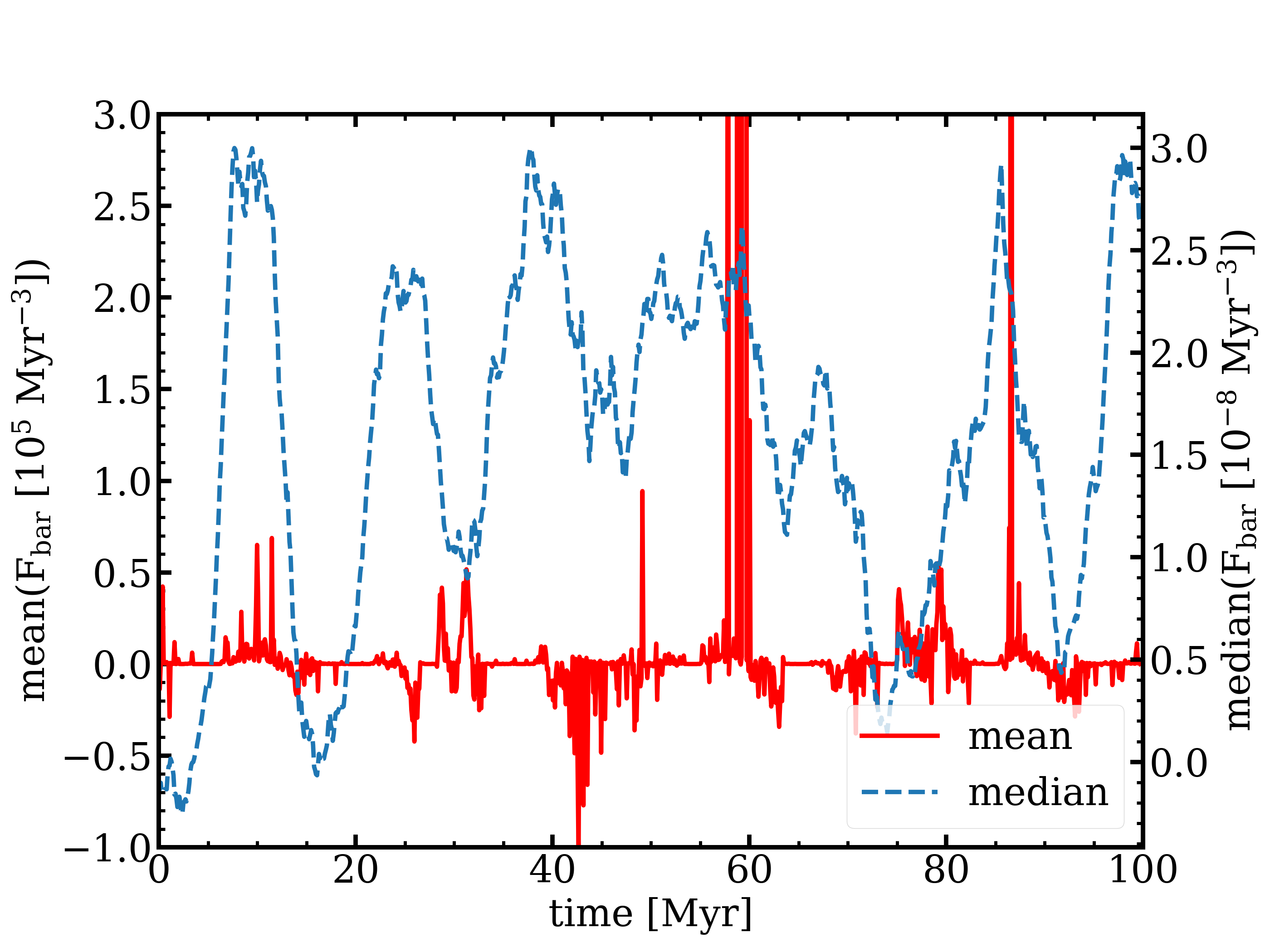}
     \vspace{-0.4cm}
     \caption{Evolution of the baroclinic term within the cluster core. The red line shows the mean, while the blue line is the median computed across the grid (notice the different left/right y-axis labelling, respectively).  To make the mean evolution more visible, we limited its plot range to $<3\times10^5 \ \Myr^{-3}$.} 
     \label{fig::mean}
     \vspace{-0.4cm}
 \end{figure}

 In Paper 1, we found that baroclinic motions are significantly weaker, and thus dynamically unimportant, than the other dynamical terms over large volumes. Yet, sporadically, the baroclinic term can become significantly large. In the following, we investigate, when, where and why, the baroclinicity increases locally in space and time.  
 
 Fig.~\ref{fig::mean} shows the temporal evolution of the mean (red) and median (blue) baroclinic term measured in the entire $(170 \ \kpc)^3$ region around the SMBH. We note that the two measurements differ by many orders of magnitude (often near 14 dex). 
 However, these large ratios are inherent to the mean which sporadically increases significantly. These huge differences between the mean and median suggest that the baroclinicity only increases in a few localized patches. 
 A comparison with the jet activity (see Figure 1 in Paper 1) shows that the periods of increasing baroclinicity are tightly correlated with the more active periods of the SMBH. 

\begin{figure*}
     \includegraphics[width = 1.01\textwidth]{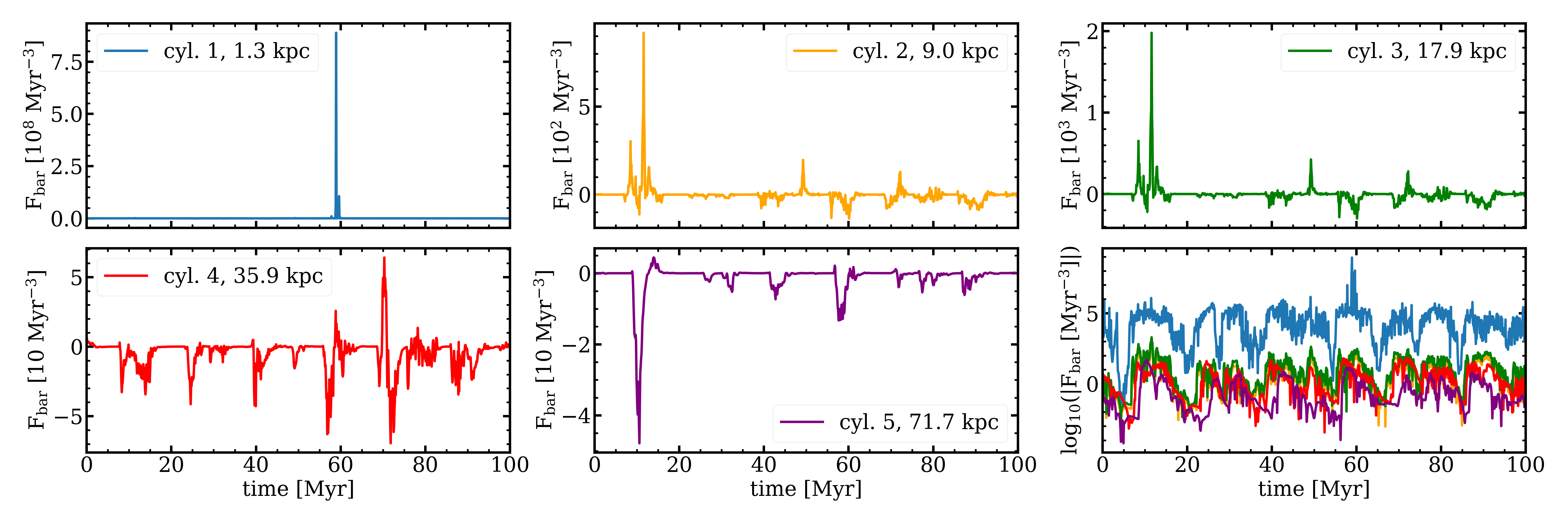}
     \vspace{-0.4cm}
     \caption{Evolution of the mean baroclinicity measured in cylindrical volumes (cyl.~1-5) aligned with the bipolar jet path and at increasing distances from the SMBH. The legend in each panel indicates the radial distances between the cylinder center and AGN (each volume has a cylindrical radius of 15 kpc; see Paper 1).
     The last panel shows the evolution of the median baroclinicity in all the considered test cylinders, showing that the inner evolution (blue line) is dominant.}
     \vspace{-0.2cm}
     \label{fig::shells}
 \end{figure*}
 
 To verify the locality of strong baroclinic motions, we measured the evolution of the mean baroclinicity inside cylindrical test volumes (see Sec. 3.1 in Paper 1 for the geometry), located at increasing distances from the SMBH and aligned with the bipolar jets. Fig.~\ref{fig::shells} shows the corresponding results. By far, baroclinicity is the strongest in the cylindrical volume that is the closest to the SMBH (cyl. 1; blue line). Here, the enstrophy is 4\,-\,5 orders of magnitude larger than in the regions at larger distances from the SMBH.
 
 Further, we measured the evolution of the median baroclinicity inside the same cylindrical test volumes, see last panel in Fig.~\ref{fig::shells}. Again, the median baroclinicity is the strongest in the cylindrical volume that is the closest to the SMBH (cyl. 1). The other four regions have comparable strength between each other, albeit always lower compared with the nuclear region ($r < 5$ kpc).
 
\begin{figure*}
  \includegraphics[width = 1.03\textwidth]{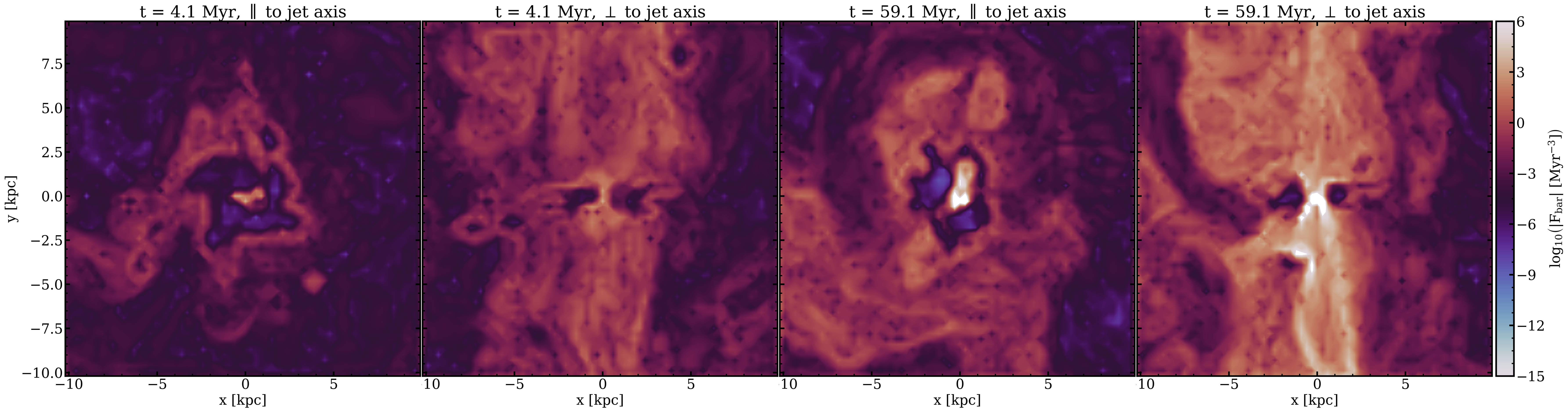}
  \vspace{-0.4cm}
  \caption{Maps of baroclinicity. Each panel shows a midplane cross-section through the simulation volume. 
  The first two panels show the baroclinic motions during the period of no AGN activity, i.e. $t = 4.1 \ \Myr$. The third and fourth panel show the baroclinic motions during the period of strong AGN activity, i.e. $t = 59.1 \ \Myr$. The first and third panels have line of sight along the jet axis. The second and fourth panel have line of sight perpendicular to the jet axis.}
  \vspace{-0.2cm}
  \label{fig::maps}
 \end{figure*}

 Fig.~\ref{fig::maps} shows midplane cross-sections of baroclinicity during either the typical quiescent or active periods. The direct inspection of these maps yields two relevant results. First, the baroclinic values are substantially weaker during the more quiescent AGN periods. Second, the major baroclinic motions are confined within both the inner SMBH region and the patches along the jet axis.

 These findings are in line with the results of Paper 1: baroclinicity and enstrophy (thus turbulence) are generated in the central region around the SMBH and are then propagated outwards along the jets. Consequently, in the next sub-section we will focus our analysis on the central $\sim (20 \ \kpc)^3$ volume around the SMBH. This volume is resolved with $\sim 300 \ \pc$. Moreover, we will focus the presentation of our results for two timesteps shown in Fig.~\ref{fig::maps}: at the most quiescent period at $4.1 \ \Myr$ and the most active period at $59.1 \ \Myr$.

%%%%%%%%%%%%%%%%%%%%%%%%%%%%%%%%%%%%%%%%%%%%%%%%%% 
%%%%%%%%%%%%%%%%%%%%%%%%%%%%%%%%%%%%%%%%%%%%%%%%%%
 \vspace{-0.5cm}
\subsection{The distribution of baroclinicity}
 
  \begin{figure*}
     \centering
     \includegraphics[width = 0.37\textwidth]{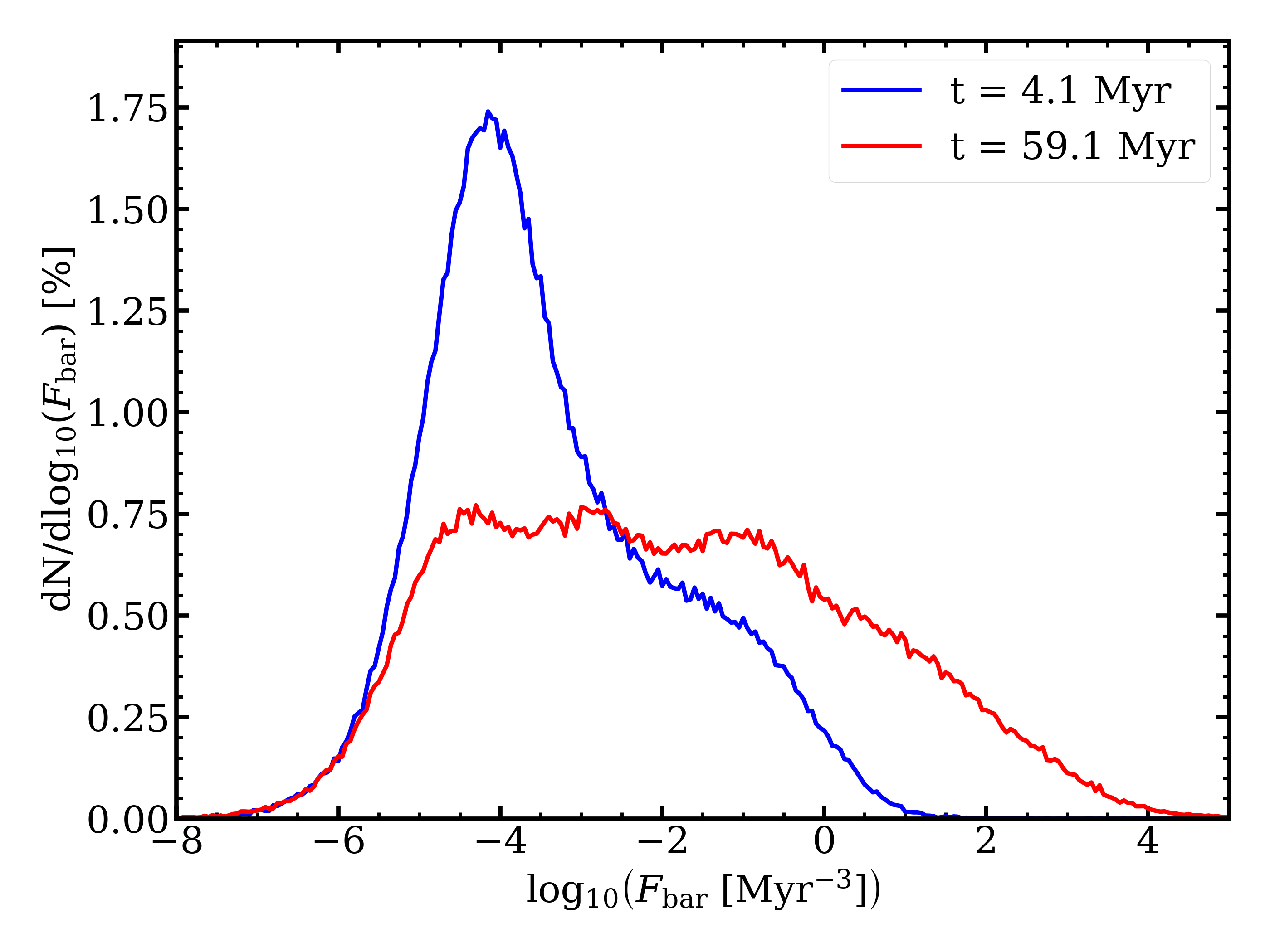} 
     \includegraphics[width = 0.37\textwidth]{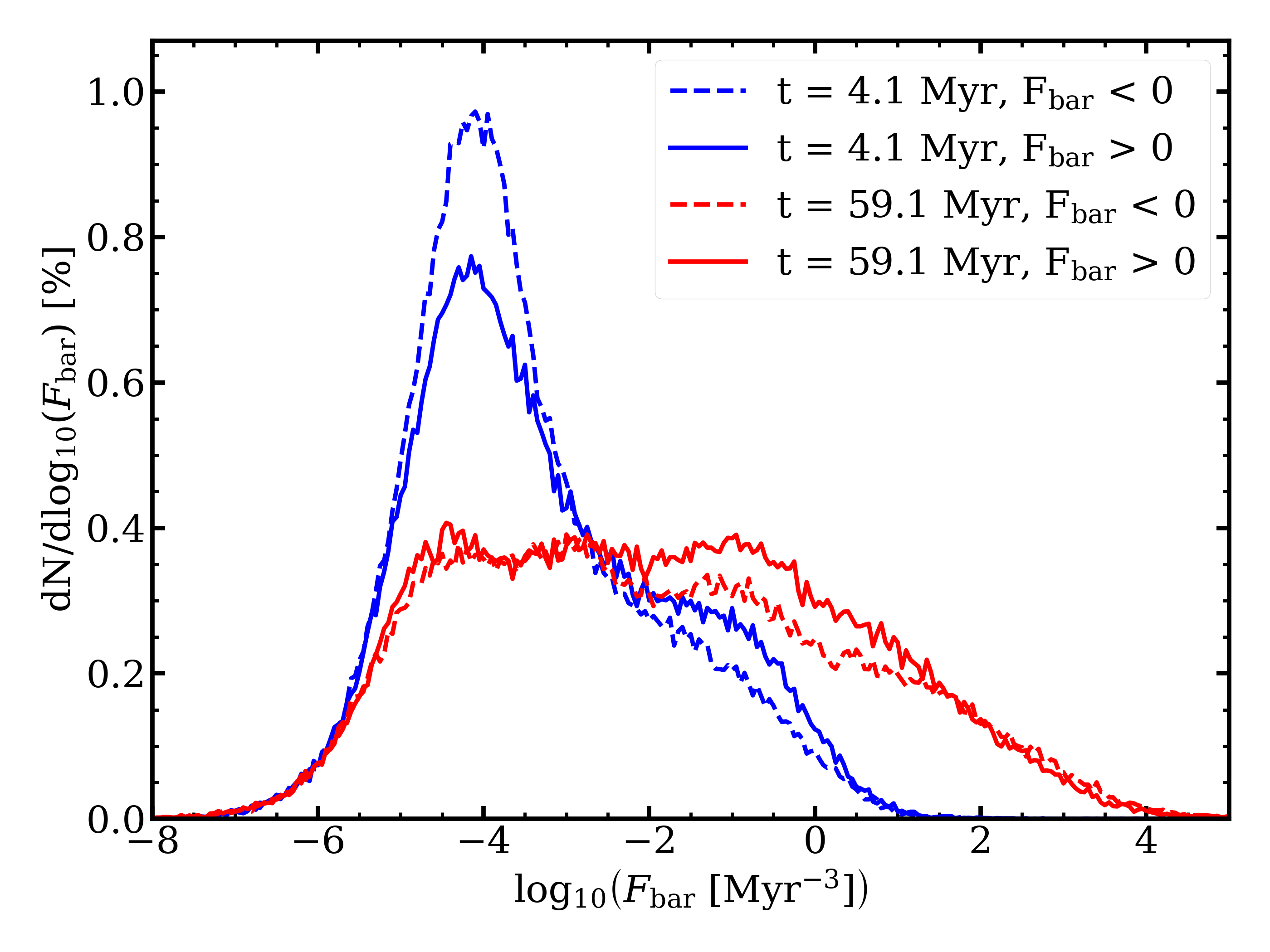} 
     \vspace{-0.4cm}
     \caption{Distribution of baroclinicity within the $(20 \ \kpc)^3$ volume around the SMBH. The left panel shows the distribution during the periods of quiescent (blue) and strong (red) AGN activity. 
     The right panel splits the two distributions where into a baroclinc source $\fb > 0$ or sink $\fb < 0$ term.} 
     \label{fig::fbar_dist}
     \vspace{-0.3cm}
 \end{figure*}
 
 In Fig.~\ref{fig::fbar_dist}, we plot the probability distribution (PDF) of baroclinicity during the two characteristic weak/strong phases ($4.1/59.1 \ \Myr$).
 In the quiescent period, the distribution peaks at $\fb \approx 10^{-4} \ \Myr^{-3}$, and it shows an asymmetric high value tail that extents up to $10 \ \Myr^{-3}$. During the active period, the distribution is instead significantly shallower, without a prominent peak. It rather shows a plateau that extends from $10^{-5} \ \Myr^{-3}$ to $10^{-1} \ \Myr^{-3}$. The high value tail extends also farther up to $10^4 \ \Myr^{-3}$. 
 These results clearly prove that baroclinic motions remain weak throughout the majority of the simulation volume, while they increase during periods of mechanical AGN feedback.
 
 We also split the PDF into parts where the baroclinic motions are positive, i.e. acting as a source $\fb > 0$, or negative, i.e. acting as a sink $\fb < 0$ (right panel in Fig.~\ref{fig::fbar_dist}). 
 During both the active and the quiescent period, baroclinic motions above $\ge 10^{-2} \ \Myr^{-3}$ are dominantly positive, i.e. acting as a source to drive turbulence in the CGM/ICM weather (alongside compressive/stretching motions at the macro scale; see Paper 1).

%%%%%%%%%%%%%%%%%%%%%%%%%%%%%%%%%%%%%%%%%%%%%%%%%%
%%%%%%%%%%%%%%%%%%%%%%%%%%%%%%%%%%%%%%%%%%%%%%%%%%
 \vspace{-0.5cm}
\subsection{The components of baroclinicity}

 \begin{figure*}
     \centering
     \includegraphics[width = 0.33\textwidth]{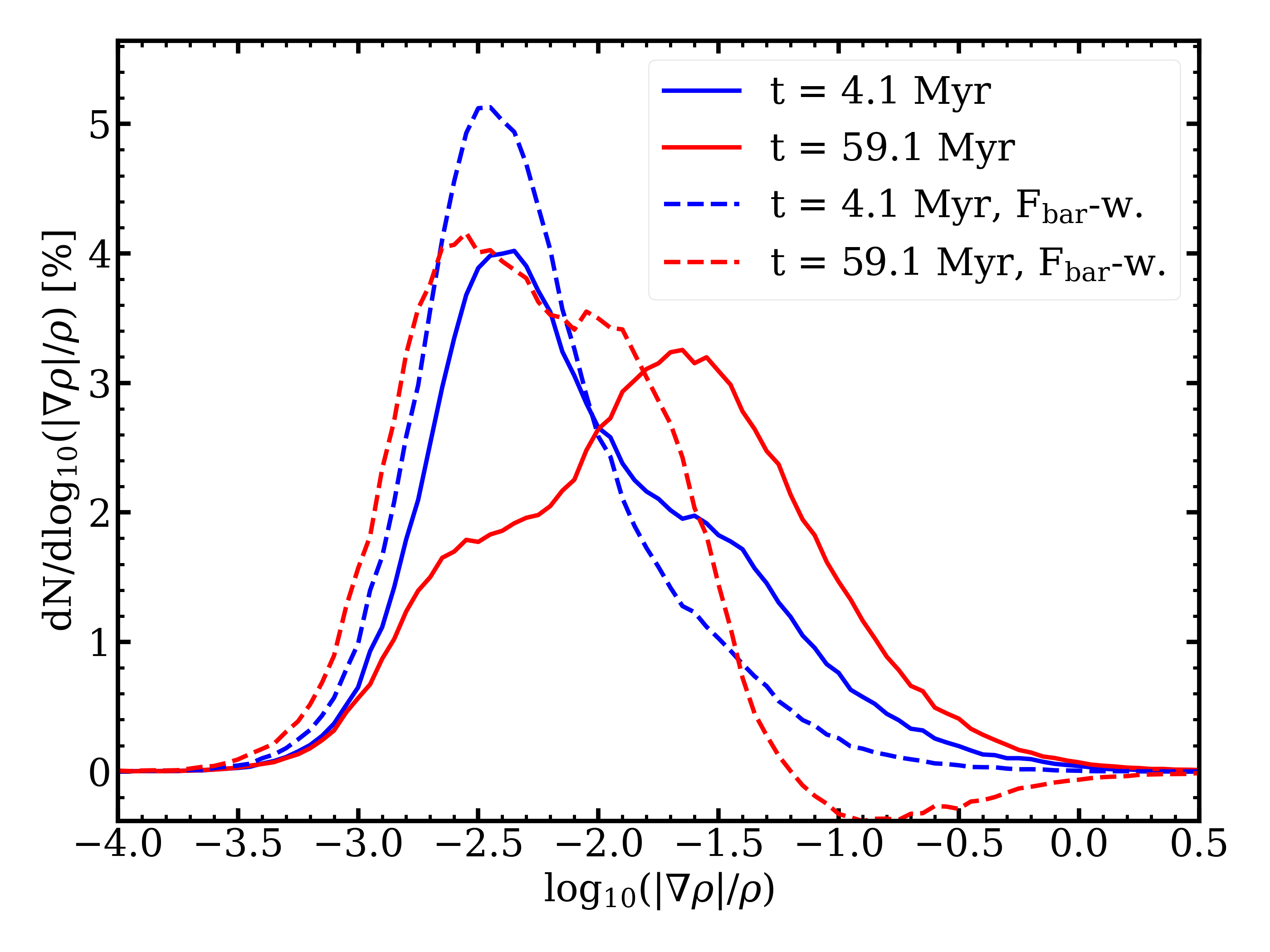} 
     \includegraphics[width = 0.33\textwidth]{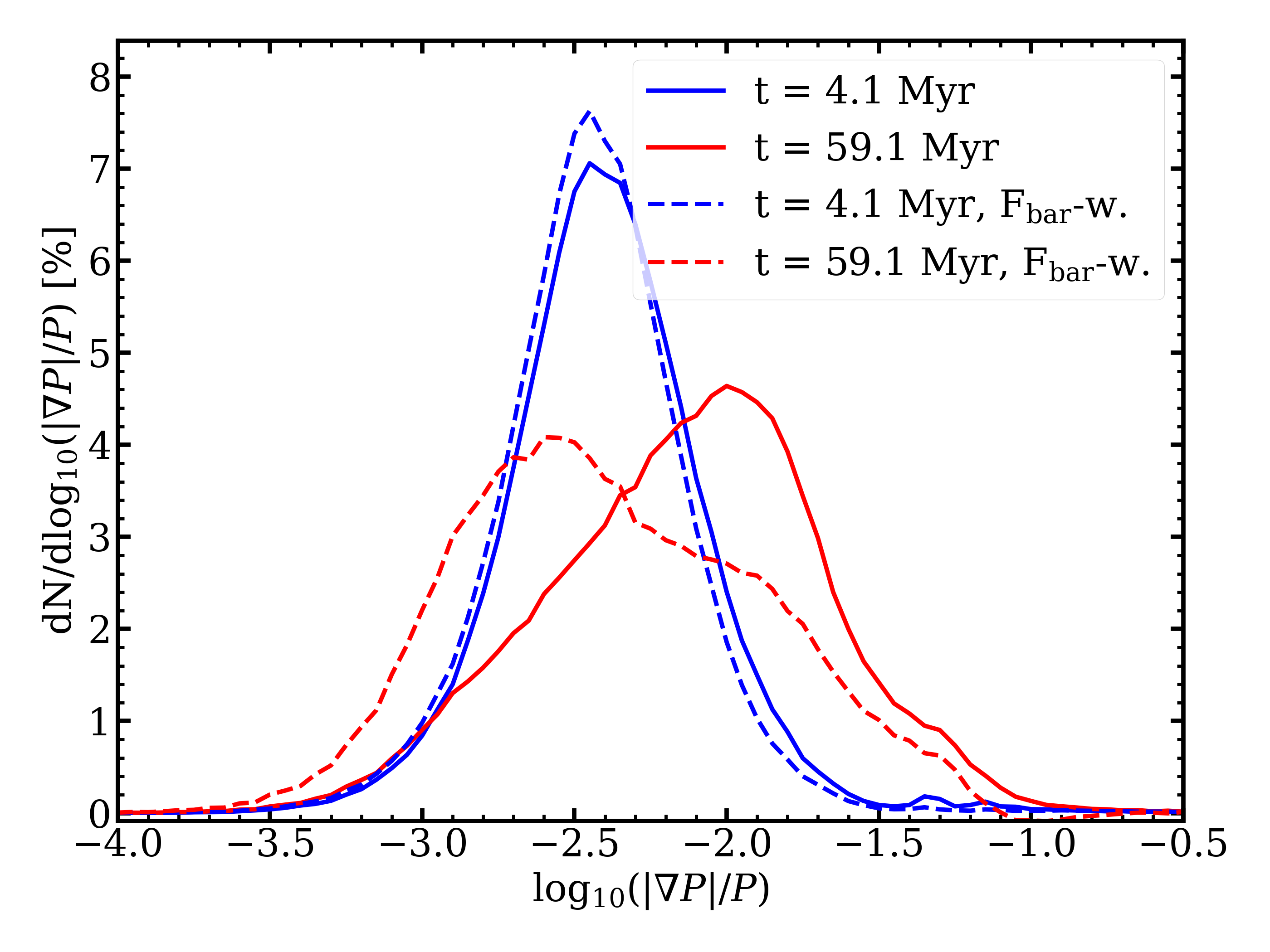} 
     \includegraphics[width = 0.33\textwidth]{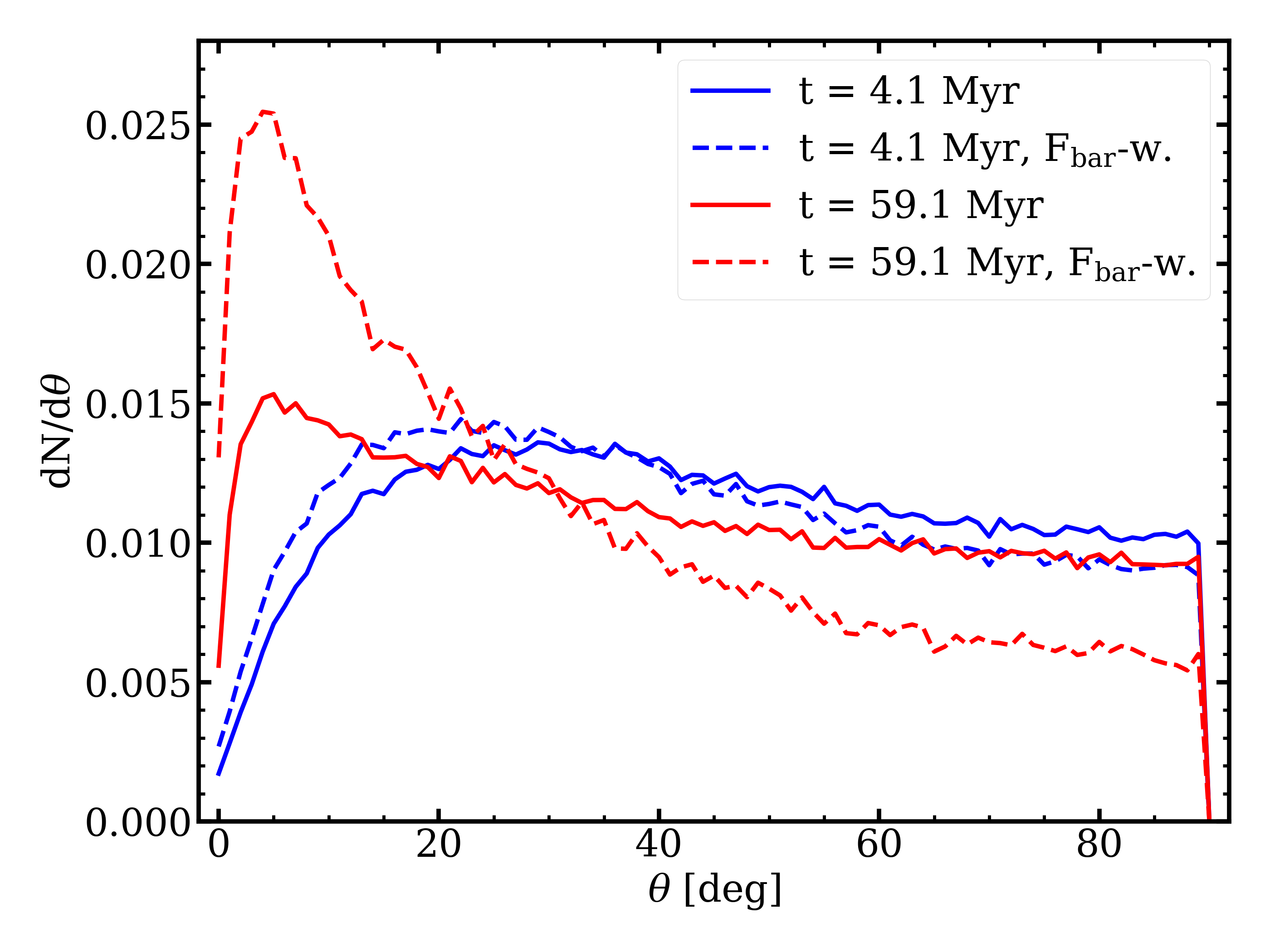} 
     \vspace{-0.4cm}
     \caption{Distribution of the three sub-terms driving baroclinicity.  
     From left to right panel: relative density gradient, relative pressure gradient, and angle between the two gradients.
     In each panel, we plot the distribution during the periods of quiescence ($t = 4.1 \ \Myr$) and strong AGN activity ($t = 59.1 \ \Myr$). Moreover, we plot the volume-weighted distribution (solid lines) and the baroclinicity-weighted distribution (dashed lines). 
     }
     \vspace{-0.2cm}
     \label{fig::dist}
 \end{figure*}

  \begin{figure*}
     \centering
     \includegraphics[width = 0.33\textwidth]{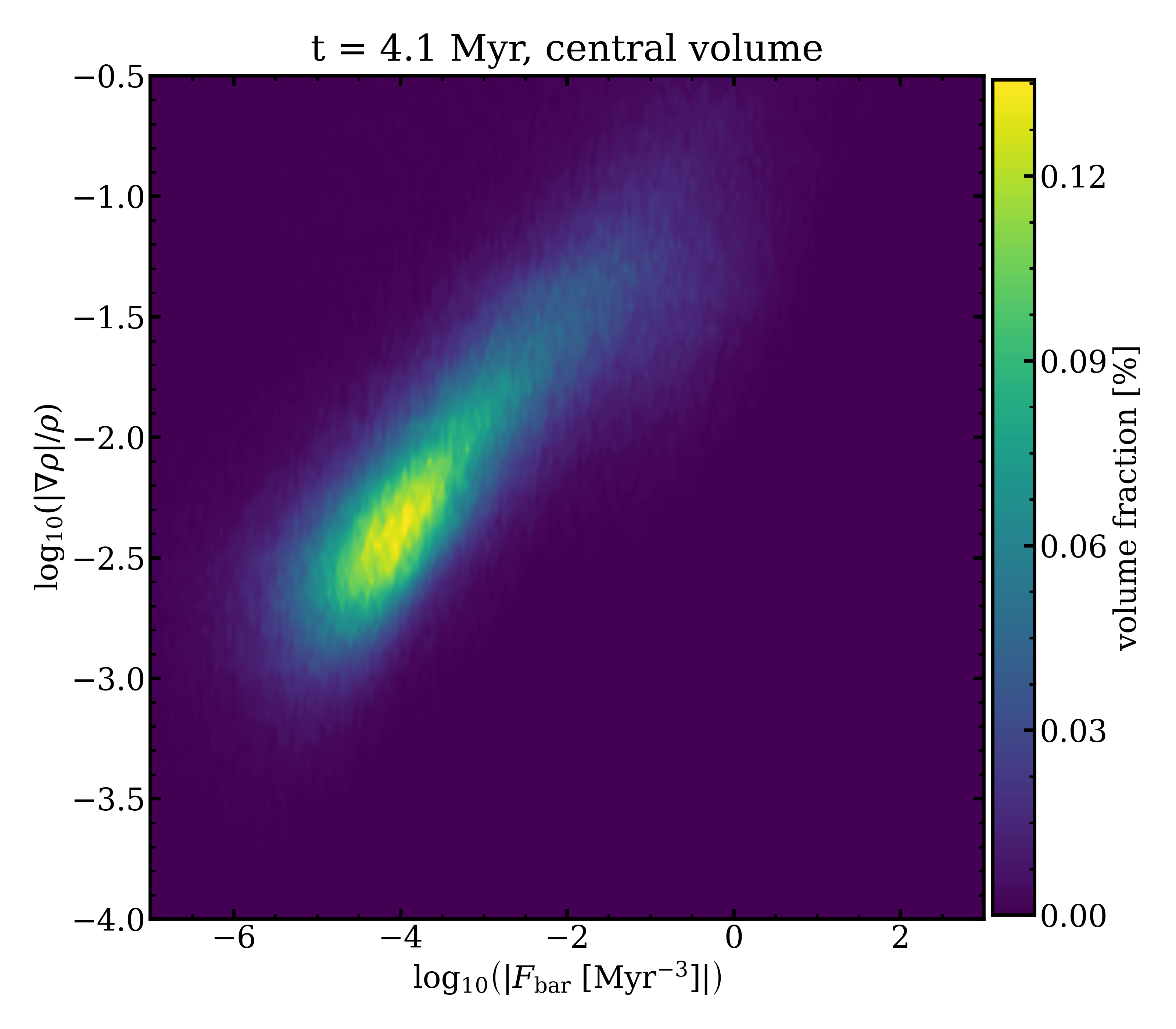}
    \includegraphics[width = 0.33\textwidth]{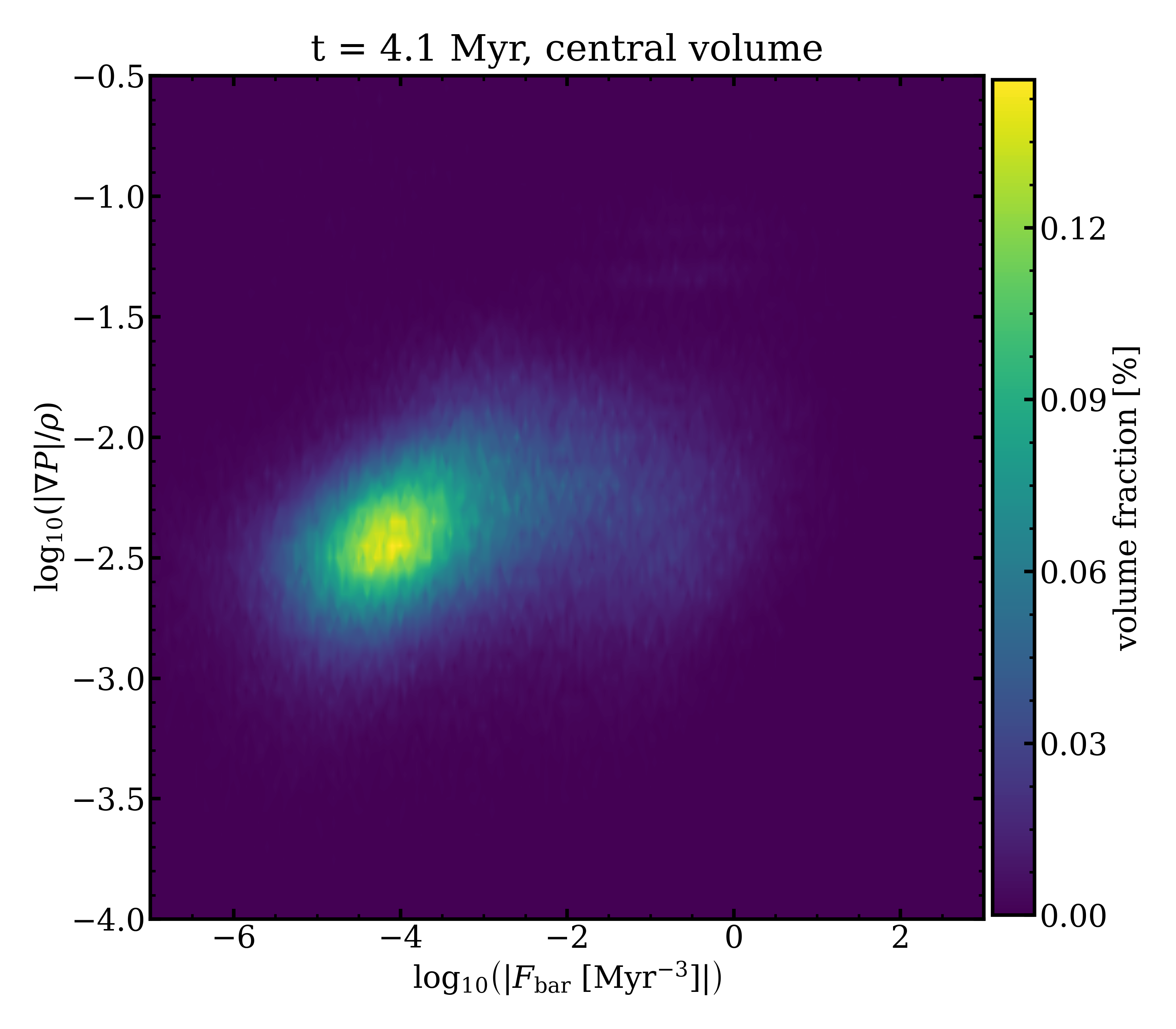}
    \includegraphics[width = 0.33\textwidth]{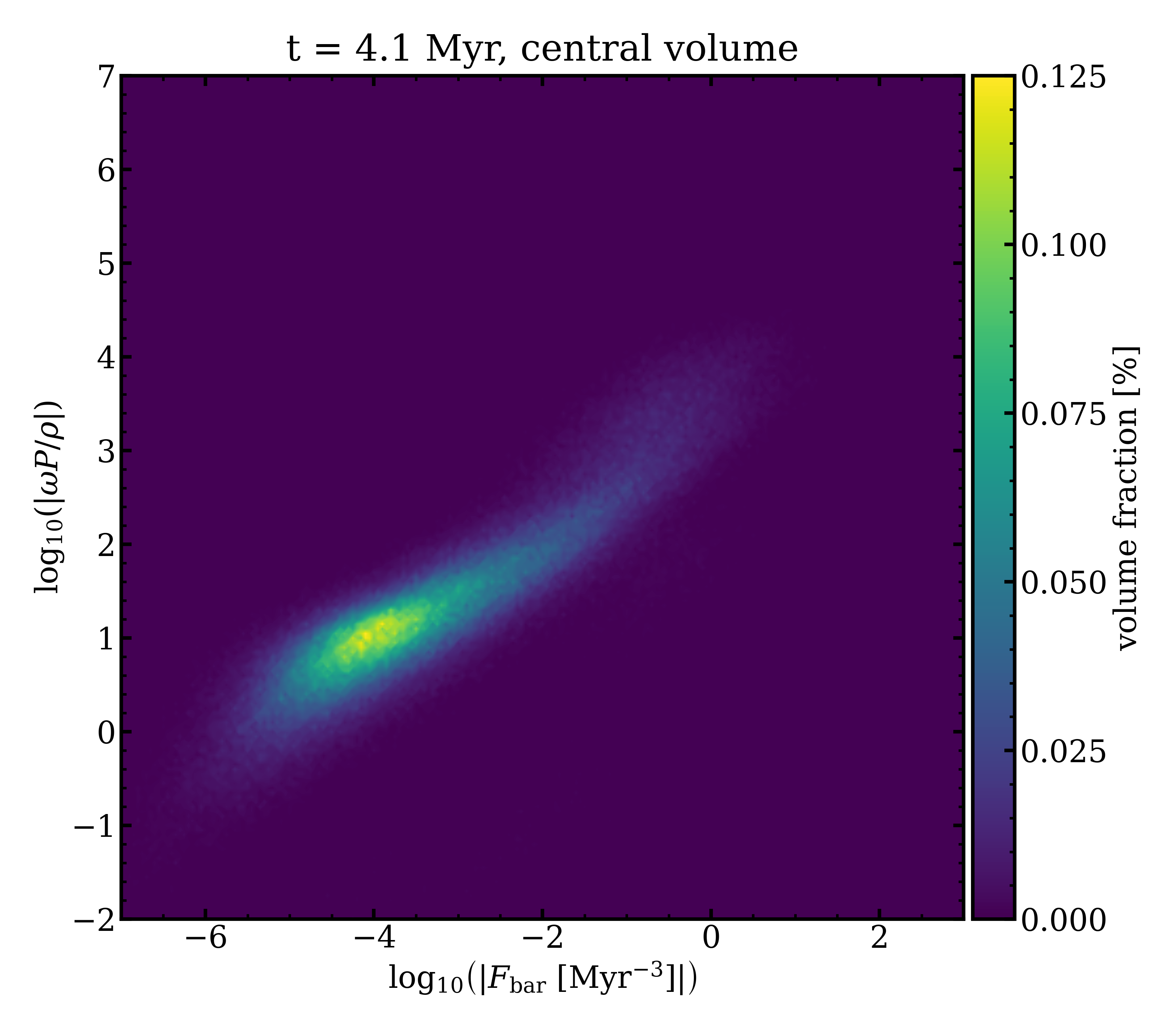} \\
     \includegraphics[width = 0.33\textwidth]{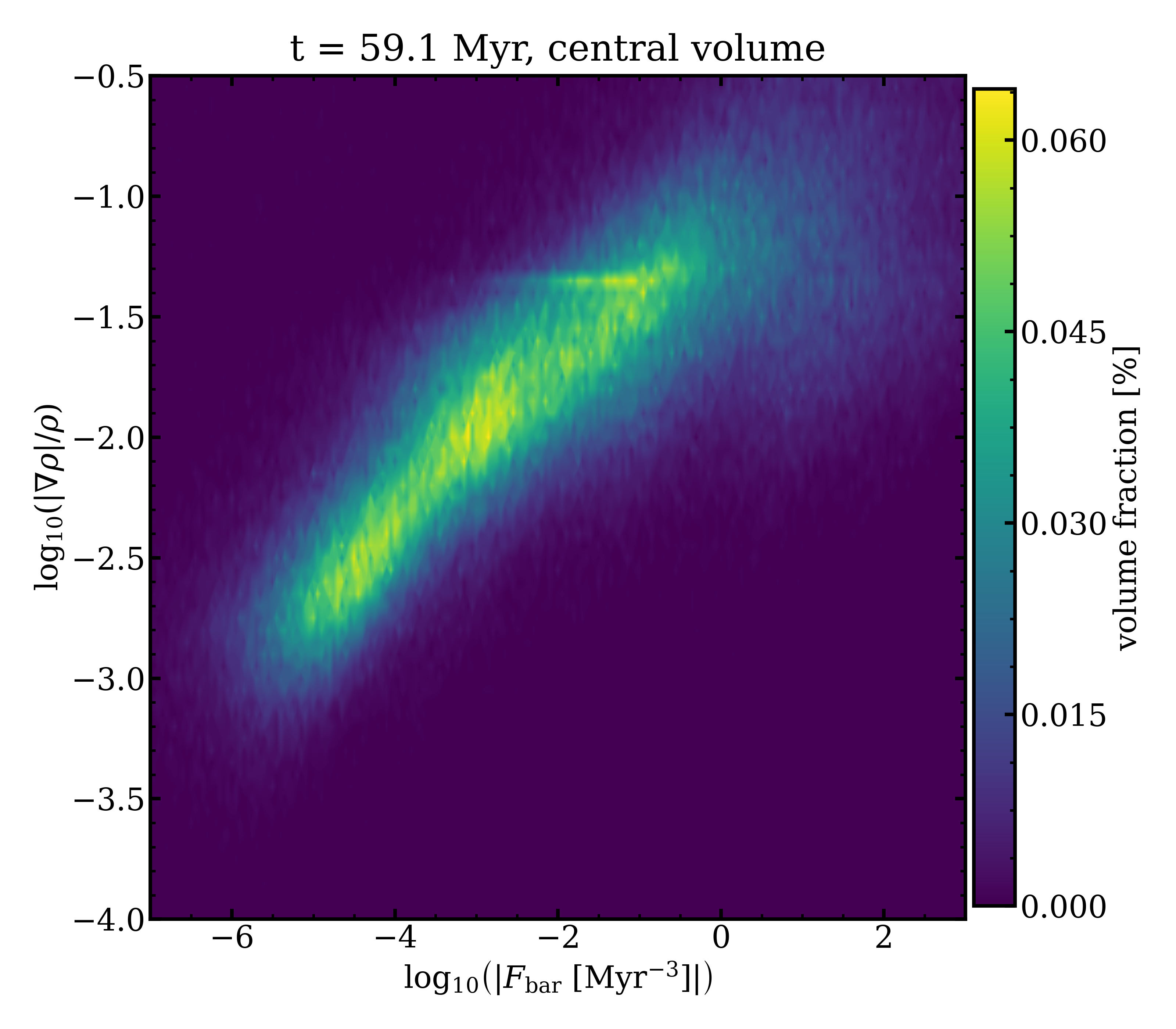}
     \includegraphics[width = 0.33\textwidth]{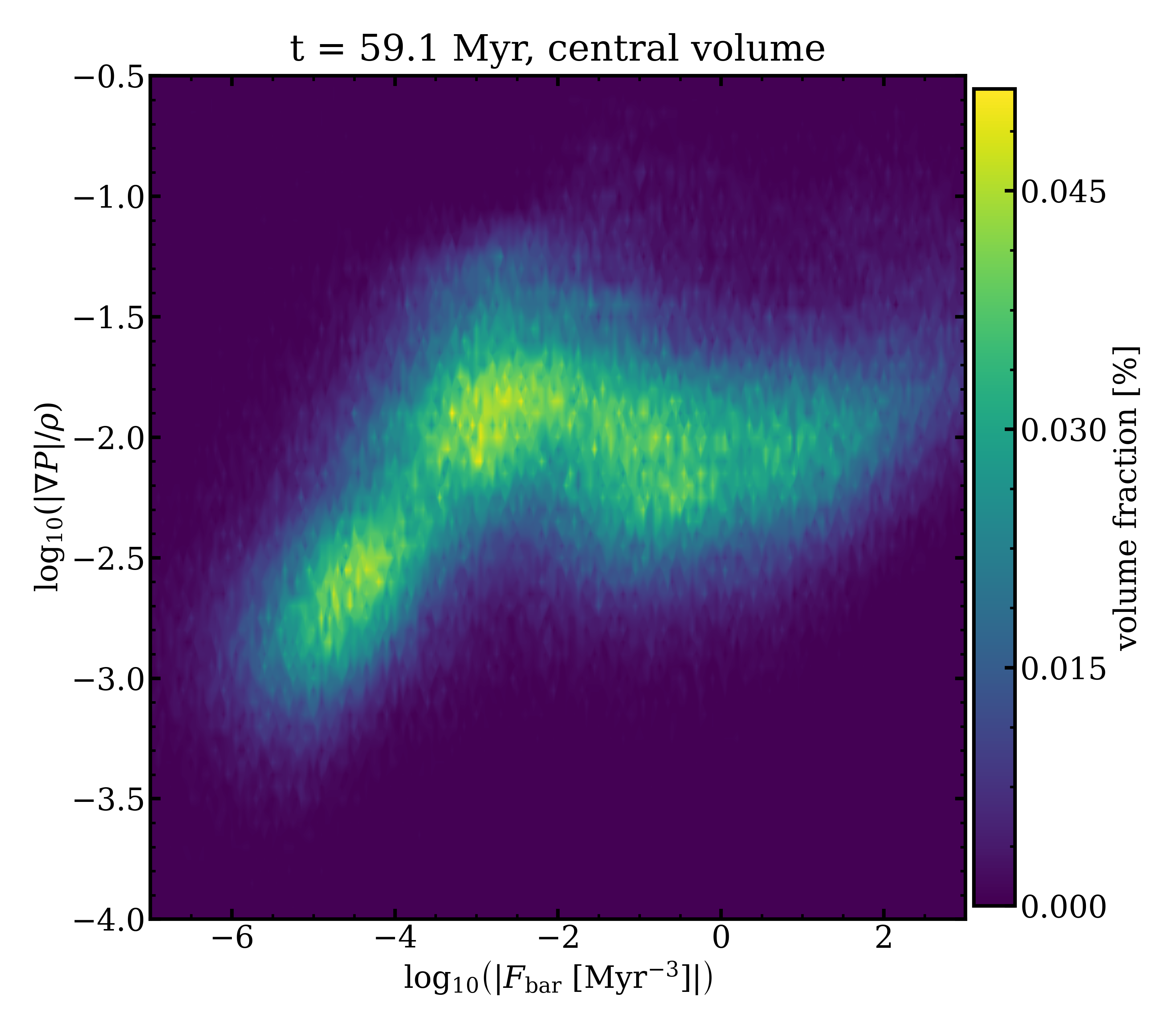} 
     \includegraphics[width = 0.33\textwidth]{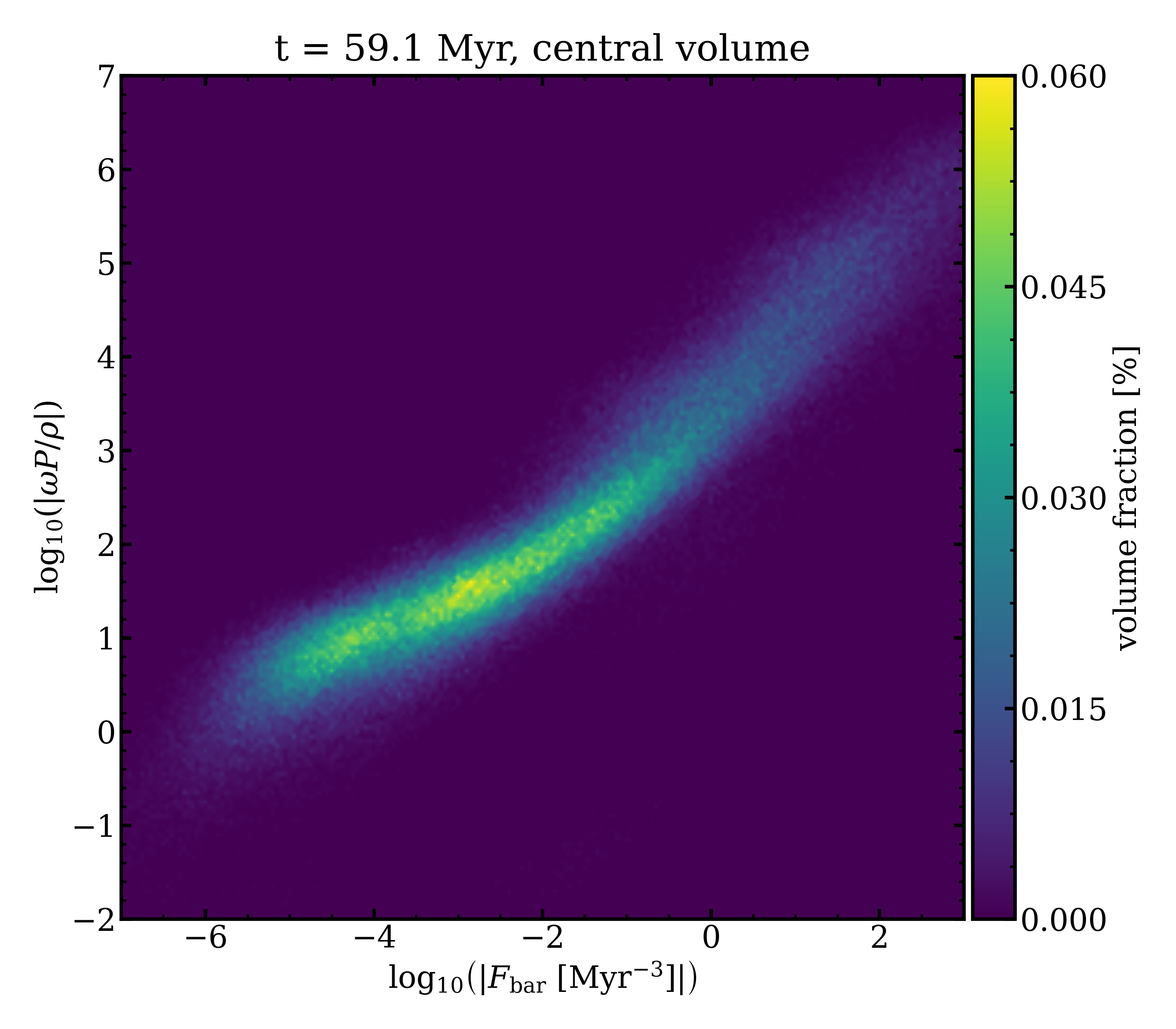}
     \caption{Correlation diagrams between the baroclinicity and the different components of baroclinicity (see Eq.~\ref{eq:Fbaro})  in the $\sim (20 \ \kpc)^3$ around the SMBH: $\nabla \rho / \rho$ (left), $\nabla P / P$ (middle), and ${\omega}\, P/\rho$ (right). The top row and bottom row show the correlation at $t = 4.1 \ \Myr$ and $t = 59.1 \ \Myr$, respectively.}
     \label{fig::phasespace2}
     \vspace{-0.4cm}
 \end{figure*} 

 In the right-hand side of Eq.~\ref{eq:Fbaro}, we have better split the main components of baroclinicity. In particular, baroclinicity is directly driven by the relative gradients of density and pressure, and the angle between the two gradients. Finally, there is a dimensional normalization that depends on vorticity and the gas temperature ($P/\rho \propto T$ for an ideal gas).  In the following, we analyse in depth each component.
 
 In Fig.~\ref{fig::dist}, we plot the volume-weighted and baroclinicity-weighted distribution of each relative gradient, as before in the central $\sim (20 \ \kpc)^3$ volume.  The first panel in Fig.~\ref{fig::dist} shows the distributions of the relative density gradient, i.e.~$\nabla \rho/\rho$. During the quiescent period, the volume-weighted and baroclinicity-weighted PDFs show a similar shape. During the period of strong AGN activity, the density gradient is significantly stronger throughout most of the volume. As a consequence, the baroclinicity-weighted distribution broadens.
  
 The second panel of Fig.~\ref{fig::dist} shows the distribution of the relative pressure gradient, i.e.~$\nabla P/P$.  These distributions tell a similar story as the distributions of the density gradient. During the quiescent period, both the volume-weighted and baroclinicity-weighted distribution show analogous shape. The pressure gradient increases during the period of AGN activity. Consequently, the baroclinicity-weighted distribution broadens as well, albeit in a more contained way compared with that of density.
 
 To further investigate the role of the gradients tied to pressure and density, we plot the correlations between each relative gradient and the baroclinic term in Fig.~\ref{fig::phasespace2}. Irrespective of the AGN activity, we observe that a clear positive correlation exists between the baroclinic term and the density gradients (first column), i.e.~the stronger the density gradient, the stronger is the baroclinicity. Yet, during the period of active feedback, the major baroclinic motions correlate even more prominently with the high-end of density gradients.
 A strong correlation is not visible for the relative pressure gradients (second column). During the quiescent periods, baroclinicity appears to be fairly independent of the pressure gradient, with a randomly scattered distribution. The lack of a tight positive correlation is also present during the active period. Indeed, the strongest baroclinic motions do not coincide with the loci of strongest pressure gradients, creating a long scattered right tail in the diagram.

 A direct comparison between the strength of the relative gradients (having simple $\rm{[1/L]}$ unit) shows that the pressure gradients tend to be on average slightly lower than the density gradients (Fig. \ref{fig::dist}).
 Such difference is however enhanced if we inspect the tails of the distribution. Indeed, in a small fraction of the volume, the relative density gradient dominates over the pressure counterpart up to $\sim$1~dex (compare first and second column in Fig.~\ref{fig::phasespace2}). 

 Vitally, baroclinic motions can only grow if the density gradient and pressure gradient are misaligned. The cross-product of the two (Eq. \ref{eq:Fbaro}) scales $\propto \sin(\theta)$, with $\theta$ being the angle between the two gradients. Here, we measured $\theta$ in the range of $\left[ 0^{\circ}, 90^{\circ}\right]$. In the right-hand panel of Fig.~\ref{fig::dist}, we plot the volume-weighted and baroclinicity-weighted distributions of $\theta$. 
 
 We find that the angle between the two gradients is not random. In the three dimensional space, the distribution of angles between two random vectors is $\propto \sin{\theta}$ \citep[e.g.][]{randomVectors}. Hence, it would peak at $90^{\circ}$ and decrease towards $0^{\circ}$. This is not the case for the distribution of $\theta$. During the quiescent period, the distributions of $\theta$ appears rather shallow with a mild peak around $20^{\circ}$ and a strong decrease towards $0^{\circ}$. This suggests that there is a preferred alignment between the two gradients. About $50 \%$ of the volume is occupied with angles below $45^{\circ}$. This is significantly larger than that for a random distribution, with $\sim33 \%$ being below $45^{\circ}$ angles. 
 
 The distribution of $\theta$ changes during the active period. Here, the distributions peaks around $5^{\circ}$. Consequently, the AGN jet feedback aligns the two gradients. Moreover, as indicated by the baroclinicity-weighted distribution, the strongest baroclinic motions are in regions where the gradients align. At first, this seems counter-intuitive as the strength of the baroclinic motions scales with $\sin{\theta}$, and $\sin(5^{\circ})$ is approximately 0.087. While this is still a small value, it is non-zero and substantial increases in the amplitude of the gradients (especially density) can compensate for the contained misalignment. 

 The last component of baroclinicity is the normalization term, ${\omega}\,P/\rho$, which combines the multiple thermodynamical units. In Fig. \ref{fig::phasespace2} (third column), we show the correlation between the normalization and the baroclinicity during the different periods. As for density, the normalization correlation stays positive during the self-regulation cycle. As feedback becomes more intense (bottom), the normalization term stretches quasi linearly over 5 dex, with a tight correlation. 
 In combination with the density gradient tails, this promotes significant amplification of the baroclinic term in the SMBH weather.

 \vspace{-0.5cm}
 \section{Conclusions} 
 
 In this work, we analysed the baroclinic motions in a simulation of recurring AGN feedback shaping the ICM/CGM atmosphere. Baroclinicity generates enstrophy and, hence, turbulence in non-barotropic and stratified atmosphere, which corresponds to misaligned pressure gradients and density gradients.  
 Here, we have analysed distribution functions of the baroclinic motions and dissected its key components: the pressure gradient, the density gradient, and the angle between the two gradients. Specifically, our analysis has focused on characteristic periods of either strong or quiescent AGN feedback activity, in particular in the $\sim (20 \ \kpc)^3$ around the SMBH.  Our results are summarized as follows:
 
 \begin{itemize}
     \item Baroclinicity is driven very locally in both time and space, being amplified during active periods of the AGN. The amplification of baroclinicity is confined within the inner region around the SMBH and the jet cocoon.
     (Fig.~\ref{fig::mean}-\ref{fig::fbar_dist})
     \item The strength of the baroclinicity depends on both the pressure and density relative gradient. 
     During active AGN periods, the distributions develop high-value tails, indicating the increase of density and pressure gradients, which increase the baroclinicity, amplifying the enstrophy. During the periods of strong AGN feedback, the baroclinic term correlates mainly with the density gradients, while there is no strong correlation with the pressure gradients. 
     (Fig.~\ref{fig::dist}-\ref{fig::phasespace2})
     \item A third ingredient of baroclinicity is the angle between the two gradients. Over different periods, the two gradients tend to be preferentially aligned, retaining small angles ($< 45^{\circ}$).  
     AGN activity drives this alignment even further, generating a clear peak at $\sim 5^{\circ}$. However, despite the contained angle, such non-zero angles can still enable significant baroclinicity in the nuclear and cocoon regions via the increased gradient amplitudes boosted by AGN jets.
     (Fig.~\ref{fig::dist})
 \end{itemize}
 
 Our work has shed new light on the evolution of baroclinicity and its role in the generation of turbulence during the cycle of AGN feeding and feedback. Baroclinic motions are subdominant compared with compressive and stretching motions over the macro scale, i.e., at large distances from the SMBH (see also Paper I).
 Yet, baroclinicity is very important at and below the meso scale around the SMBH ($r<10$ kpc) during periods of strong AGN jet activity.
 These findings elucidate the evolution of turbulence during periods of AGN feedback: primordial enstrophy, hence subsonic turbulence, is generated close to the SMBH (within the BCG) and is then advected outwards along the jet axis and cocoon. During this cycle, baroclinicity is mainly relevant for the \textit{seeding} of enstrophy close to the the SMBH. On large spatial ($r>10$ kpc) and temporal scales ($t>20$ Myr), baroclinicity is subdominant for the evolution of enstrophy/turbulence. On such macro scales, compression and stretching motions mainly amplify the evolution of enstrophy. However, neither of these two modes are able to generate enstrophy out of a zero $\epsilon$ floor. 
 Our findings thus indicate that baroclinicity is not affecting the macro evolution but is very relevant for the initial seeding of fresh turbulence close to the SMBH.
 
 In closing, our study provides a novel step forward in understanding astrophysical atmospheres toward a unified \texttt{BlackHoleWeather} framework, akin to the complexity of Earth's weather. Indeed, albeit the nature of the underlying baroclinic instability differs in the details (Sec.~\ref{s:intro}), a broad analogy between these spectacular astrophysical events and `cyclonic storms' in terrestrial meteorology exists due to the key role of baroclinicity in both gas-dynamical evolutions.

\vspace{-0.5cm}
\section*{Data availability}

The data underlying this article will be shared on reasonable request to the authors, unless being in conflict or breaking the privacy of ongoing related work led by our collaboration members.

 \vspace{-0.5cm}
 \section*{Acknowledgements}
 D.W.~is funded by DFG (Deutsche Forschungsgemeinschaft) 441694982. 
 M.G.~acknowledges partial support by HST GO-15890.020/023-A and \texttt{BlackHoleWeather} program.
 HPC resources were in part provided by the NASA/Pleiades HEC Program (SMD-1726).
 We are thankful for the recent workshop \href{https://www.sexten-cfa.eu/event/multiphase-agn-feeding-feedback-ii-linking-the-micro-to-macro-scales-in-galaxies-groups-and-clusters}{`Multiphase AGN Feeding \& Feedback II'}
 (held at SCfA, Italy) which has enabled key fruitful discussions.  
\vspace{-0.5cm}
 \bibliographystyle{biblio.bst}
 \bibliography{mybib}
  
\label{lastpage}
\end{document}